\documentclass[11pt]{article}

\usepackage{amsmath,amsfonts,amssymb,caption,graphicx,hypernat}
\usepackage{hyperref}
\hypersetup{colorlinks=true,pdftitle="Sumset inequalities for differential entropy",pdftex}

\newcommand{\X}{{\rm X}}
\newcommand{\Y}{{\rm Y}}
\newcommand{\Z}{{\rm Z}}

\newcommand{\RL}{{\mathbb R}}

\newcommand{\dist}{\mbox{\rm dist}}

\newcommand{\VAR}{\mbox{\rm Var}}

\def\ba{\begin{align}}
\def\ea{\end{align}}
\def\ban{\begin{align*}}
\def\ean{\end{align*}}

\def\be{\begin{eqnarray}}
\def\ee{\end{eqnarray}}
\def\ben{\begin{eqnarray*}}
\def\een{\end{eqnarray*}}

\def\elabel#1{\label{e:#1}}

\def\sq{$\Box$}

\def\qed{\ifmmode\sq\else{\unskip\nobreak\hfil
\penalty50\hskip1em\null\nobreak\hfil\sq
\parfillskip=0pt\finalhyphendemerits=0\endgraf}\fi\par\medbreak}

\newsavebox{\junk}
\savebox{\junk}[1.6mm]{\hbox{$|\!|\!|$}}

\def\det{{\mathop{\rm det}}}

\def\til={{\widetilde =}}

\def\half{{\mathchoice{\textstyle \frac{1}{2}}%
{\frac{1}{2}}%
{\hbox{\tiny $\frac{1}{2}$}}%
{\hbox{\tiny $\frac{1}{2}$}} }}

 \def\eq#1/{(\ref{#1})}

\newtheorem{theorem}{Theorem}[section]
\newtheorem{corollary}[theorem]{Corollary}
\newtheorem{proposition}[theorem]{Proposition}
\newtheorem{lemma}[theorem]{Lemma}

\def\eq#1/{(\ref{e:#1})}

\newcommand{\beqn}[1]{\notes{#1}%
\begin{eqnarray} \elabel{#1}}

\newcommand{\eeqn}{\end{eqnarray} }

\newcommand{\beq}[1]{\notes{#1}%
\begin{equation}\elabel{#1}}

\newcommand{\eeq}{\end{equation}} 

\def\bdes{\begin{description}}
\def\edes{\end{description}}

\def\notes#1{}

\begin{document}

\title{\vspace{-1.5cm}%
Sumset and Inverse Sumset Inequalities\\
for Differential Entropy and Mutual Information}

\author
{
        Ioannis Kontoyiannis, {\sl Fellow, IEEE}
    \thanks{Department of Informatics,
        Athens University of Economics and Business,
        Patission 76, Athens 10434, Greece.
                Email: {\tt yiannis@aueb.gr}.
	}
    \thanks{I.K.\ was supported, in part, by a Marie Curie 
	International Outgoing Fellowship, PIOF-GA-2009-235837.
        }
\and
	Mokshay Madiman, {\sl Member, IEEE}
    \thanks{Department of Statistics, 
	Yale University, 24 Hillhouse Avenue, 
	New Haven, CT 06511, USA.
		Email: {\tt mokshay.madiman@yale.edu}.
	}
	\thanks{M.M.\ was supported by the NSF CAREER grant 
	DMS-1056996 and by NSF grant CCF-1065494.
	}
}

\footnotetext{Preliminary versions of parts of this work
were presented at the 2008 IEEE Information Theory Workshop
\cite{madiman-itw:08} and at 
the 2010 IEEE International Symposium on Information Theory 
\cite{kontoyiannis-MM-ISIT:10}.}

\date{\today}

\maketitle

\begin{abstract}
The {\em sumset} and {\em inverse sumset} 
theories of Freiman, Pl\"{u}nnecke and Ruzsa,
give bounds connecting the cardinality 
of the sumset $A+B=\{a+b\;;\;a\in A,\,b\in B\}$
of two discrete sets $A,B$, to the 
cardinalities (or the finer structure) 
of the original sets $A,B$.
For example, the sum-difference bound
of Ruzsa states that, 
$|A+B|\,|A|\,|B|\leq|A-B|^3$,
where the difference set 
$A-B= \{a-b\;;\;a\in A,\,b\in B\}$.
Interpreting the differential entropy $h(X)$
of a continuous random variable $X$ as 
(the logarithm of) the size of 
the effective support
of $X$, the main contribution of this paper
is a series of natural information-theoretic 
analogs for these results.
For example, the Ruzsa sum-difference bound
becomes the new inequality,
$h(X+Y)+h(X)+h(Y)\leq 3h(X-Y)$, for any 
pair of independent continuous random variables
$X$ and $Y$. Our results include 
differential-entropy versions 
of Ruzsa's triangle inequality,
the Pl\"{u}nnecke-Ruzsa inequality,
and the Balog-Szemer\'{e}di-Gowers lemma.
Also we give a differential entropy version 
of the Freiman-Green-Ruzsa inverse-sumset theorem,
which can be seen as a quantitative converse 
to the entropy power inequality.
Versions of most of these results
for the discrete entropy $H(X)$ were recently
proved by Tao, relying heavily 
on a strong, functional form of 
the submodularity property of $H(X)$.
Since differential entropy is {\em not} 
functionally submodular, 
in the continuous case 
many of the corresponding discrete proofs fail,
in many cases requiring substantially
new proof strategies. 
We find that the basic property that 
naturally replaces the discrete 
functional submodularity, is the data 
processing property of mutual 
information.
\end{abstract}

{\small

\noindent
{\bf Keywords ---}
Shannon entropy, differential entropy, sumset bounds,
inequalities, submodularity, data processing, mutual information
}

\newpage

\section{Introduction}

\subsection{Motivation}

Roughly speaking, 
the field of {\em additive combinatorics}
provides tools that allow us to count
the number of occurrences of particular additive structures 
in specific subsets of a discrete group;
see \cite{TV06:book} for a broad introduction.
The prototypical example is the study of 
the existence of arithmetic progressions 
within specific sets of integers --
as opposed to the multiplicative structure 
that underlies prime factorization and much of 
classical combinatorics and number theory. 
There have been
several major developments and a lot of high-profile mathematical
activity in connection with
additive combinatorics in recent years, perhaps
the most famous example being the celebrated Green-Tao 
theorem on the existence of arbitrarily long
arithmetic progressions within the set
of prime numbers.

An important collection of tools in
additive combinatorics is a variety of 
{\em sumset inequalities}, the so-called
Pl\"{u}nnecke-Ruzsa {\em sumset theory}; 
see \cite{TV06:book} for details.
The 
{\em sumset}
$A+B$ of two discrete sets $A$ and $B$ 
is defined as,
$A+B=\{a+b: a\in A, b\in B\}$,
and a {\em sumset inequality}
is an inequality connecting the cardinality
$|A+B|$ of $A+B$ with
the cardinalities $|A|,|B|$ of $A$ and $B$,
respectively.
For example, there are the obvious bounds,
\be
\max\{|A|,|B|\}\leq|A+B|\leq|A|\,|B|,
\label{eq:Itrivial}
\ee
as well as much more subtle results, like
the Ruzsa triangle inequality \cite{ruzsa:96},
\be
|A-C|\leq \frac{|A-B|\,|B-C|}{|B|},
\label{eq:Itriangle}
\ee
or the sum-difference bound \cite{ruzsa:96},
\be
|A+B| \leq\frac{|A-B|^3}{ |A|\,|B| },
\label{eq:Isum-difference}
\ee
all of which hold for arbitrary subsets
$A,B,C$ of the integers or any other 
discrete abelian group,
and where the
{\em difference set}
$A-B$ is defined as,
$A-B=\{a-b: a\in A, b\in B\}$.

In the converse direction, the
Freiman-Ruzsa {\em inverse sumset theory}
provides information about sets $A$
for which $|A+A|$ is close to being
as small as possible; see 
Section~\ref{s:inverse-h}
for a brief discussion or the text
\cite{TV06:book} for details.

In this context, recall that
Shannon's asymptotic equipartition 
property (AEP) \cite{cover:book}
says that the entropy $H(X)$ of a discrete 
random variable $X$
can be thought of as the logarithm 
of the {\em effective cardinality}
of the alphabet of $X$. This suggests 
a correspondence between bounds
for the cardinalities of sumsets,
e.g., $|A+B|$, and corresponding bounds
for the entropy of sums of independent
discrete 
random variables, e.g., $H(X+Y)$.
First identified by Ruzsa \cite{Ruz09}, 
this connection
has also been explored
in the last few years 
in different directions by,
among others,
Tao and Vu \cite{tao-vu:notes},
Lapidoth and Pete \cite{lapidoth-pete},
Madiman and Kontoyiannis \cite{kontoyiannis-MM-ISIT:10},
and 
Madiman, Marcus and Tetali \cite{MMT08:pre};
additional pointers to the relevant literature
are given below.

This connection was developed most extensively
by Tao in \cite{tao:10}. 
The main idea is to replace sets
by (independent, discrete) random variables, 
and  then replace the
log-cardinality, $\log |A|$, of each set $A$
by the (discrete, Shannon) entropy
of the corresponding random variable
(where $\log$ denotes the natural 
logarithm $\log_e$).
Thus,
for independent discrete random variables
$X,Y,Z$,
the simple bounds~(\ref{eq:Itrivial}) become,
\ben
\max\{H(X),H(Y)\}\leq H(X+Y)\leq H(X)+H(Y),
\een
which is a trivial exercise in manipulating entropy \cite{cover:book}.
On the other hand, again for independent discrete random variables,
Ruzsa's influential bounds~(\ref{eq:Itriangle})
and~(\ref{eq:Isum-difference}) become,
respectively,
\ben
H(X-Z)+H(Y)&\leq& H(X-Y)+H(Y-Z)\\
\mbox{and}\;\;
H(X+Y)+H(X)+H(Y)&\leq& 3 H(X-Y),
\een
which are nontrivial facts proved in \cite{tao:10}.

Our main motivation is to
examine the extent to which this 
analogy can be carried further:
According to the AEP \cite{cover:book},
the {\em differential entropy} $h(X)$ of a
{\em continuous} random variable $X$ can be thought
of as the logarithm of the 
``size of the effective support'' of $X$.
In this work we state and prove
natural ``differential entropy analogs''
of various sumset and inverse-sumset 
bounds, many of which were proved 
for the discrete Shannon entropy
in the recent work of Tao \cite{tao:10}
and in earlier papers by
Kaimonovich and Vershik \cite{KV:83},
Tao and Vu \cite{tao-vu:notes},
Madiman \cite{madiman-itw:08},
Ruzsa \cite{Ruz09}, 
and Madiman, Marcus and Tetali \cite{MMT08:pre}.

Of particular interest in motivating these 
results is the fact that
the main technical ingredient in the proofs
of many of the corresponding discrete bounds
was a strong, functional
form of the {\em submodularity} property
of the discrete Shannon entropy;
see Section~\ref{s:prelim}
for details. The fact that 
differential entropy is {\em not}
functionally submodular was the source of the 
main difficulty as well as the main interest
for the present development.

\subsection{Outline of main results}

In Section~\ref{s:prelim}, after briefly reviewing
some necessary background and basic 
definitions,
we discuss the functional submodularity
of the discrete entropy, and explain
how it fails for differential entropy.

Section~\ref{s:sumset-h} contains most of our main
results, namely, a series of natural differential entropy
analogs of the sumset bounds in \cite{tao:10}
and in the earlier papers mentioned above.
In Theorem~\ref{thm:Ctriangle} we prove the following
version of the {\em Ruzsa triangle inequality}:
If $X,Y,Z$ are independent, then:
$$h(X-Z)\leq h(X-Y)+h(Y-Z)-h(Y).$$
In Theorem~\ref{thm:Cdoublediff}, we prove
the {\em doubling-difference inequality}:
If $X_1,X_2$ are 
independent and identically distributed 
(i.i.d.), then:
$$
\frac{1}{2}
\leq
\frac{h(X_1+X_2)-h(X_1)}{h(X_1-X_2)-h(X_1)}
\leq 2.
$$
More generally, when $X_1,X_2$ are independent but not
identically distributed, the following
{\em sum-difference inequality} holds,
given in Theorem~\ref{thm:sum-diff}:
If $X_1,X_2$ are  independent, then:
$$h(X_1+X_2)\leq 3h(X_1-X_2)-h(X_1)-h(X_2).$$

A version of the
{\em Pl\"{u}nnecke-Ruzsa inequality} 
for differential entropy is given
in Theorem~\ref{thm:CPR}:
If $X,Y_1,Y_2,\ldots,Y_n$ are independent
and there are constants $K_1,K_2,\ldots,K_n$
such that,
$$h(X+Y_i)\leq h(X) + \log K_i,\;\;\;\;\mbox{for each }i,$$
then
$$h(X+Y_1+Y_2+\cdots+Y_n)\leq h(X)+\log K_1K_2\cdots K_n.$$
An application of this result gives
the {\em iterated sum bound} of Theorem~\ref{thm:iterated}.

Next we prove a {\em Balog-Szemer\'{e}di-Gowers lemma}
for differential entropy. It says that, if $X,Y$ are 
weakly dependent and $X+Y$ has small entropy, 
then there exist conditionally independent 
versions of $X,Y$
that have almost the same entropy, and 
whose {\em independent} sum still has small entropy.
Specifically, in Theorem~\ref{thm:CBSG} we prove
the following: Suppose $X,Y$ satisfy
$I(X;Y)\leq \log K$, 
for some $K \geq 1$, and suppose also that, 
$$ h(X+Y) \leq \frac{1}{2} h(X) + \frac{1}{2} h(Y) + \log K.$$
Let $X_1,X_2$ be conditionally independent versions of $X$ given $Y$,
and let $Y'$ be a conditionally independent version of $Y$,
given $X_2$ and $Y$. 
Then:
\ben
h( X_2 | X_1, Y ) 	&\geq& h(X) - \log K \\
h( Y' | X_1, Y ) 	&\geq& h(Y) - \log K \\
h( X_2 + Y' | X_1, Y )  &\leq& \frac{1}{2} h(X) 
+ \frac{1}{2} h(Y) + 7 \log K.
\een

The main interest in the proofs of the above results 
is that, in most cases, the corresponding discrete-entropy 
proofs do not generalize in a straightforward manner.
There, the main technical tool used
is a strong, functional submodularity property 
of $H(X)$, which does {\em not} hold for differential 
entropy. Moreover, for several results
it is the overall proof structure 
that does not carry over to the 
continuous case;
not only the method, but some of the
important intermediate steps fail 
to hold for differential entropy,
requiring substantially 
new proof strategies. 

The main technical ingredient in our proofs
is the {\em data processing}
property of mutual information.
Indeed, most of the bounds in Section~\ref{s:sumset-h}
can be equivalently stated in terms of mutual
information instead of differential entropy.
And since data processing is universal
in that it holds regardless of the space 
in which the relevant random variables take values,
these proofs offer alternative derivations
for the discrete counterparts of the results.
The earlier discrete versions are 
discussed in Section~\ref{s:discrete},
where we also describe the entropy version
of the {\em Ruzsa covering lemma} and the fact
that its obvious generalization fails in 
continuous case.

In Section~\ref{s:inverse-h} we give
a version of the Freiman-Green-Ruzsa 
inverse-sumset theorem for differential entropy.
Roughly speaking, Tao in \cite{tao:10} proves that, 
if the entropy $H(X+X')$ of the sum of two i.i.d.\ copies
of a discrete random variable $X$ is close to $H(X)$,
then $X$ is approximately uniformly distributed on
a generalized arithmetic progression.

In the continuous case, the entropy power inequality
\cite{cover:book} says that, if $X,X'$ are i.i.d.,
then,
$$h(X+X')-h(X)\geq\frac{1}{2}\log 2,$$
with equality if and only if $X$ is Gaussian.
In Theorem~\ref{thm:inverse} we state and prove
a quantitative converse to this statement: Under
certain regularity conditions on the
density of $X$, we show that if $h(X+X')$
is not much larger than $h(X)$, then 
$X$ will necessarily be approximately Gaussian,
in that the relative entropy between its density 
and that of a Gaussian with the same
mean and variance will be correspondingly
small.

Finally we note that,
in view of the fact that additive noise is one
of the most common modeling assumptions in 
Shannon theory, it is natural to expect that,
likely, some of the bounds developed here may
have applications in core information-theoretic
problems. Preliminary connections in this
direction can be found in the recent work
of Cohen and Zamir \cite{cohen-zamir:08},
Etkin and Ordentlich \cite{etkin-eor:09},
and Wu, Shamai and Verd\'{u} \cite{WVS:pre}.

\section{Elementary Bounds and Preliminaries}
\label{s:prelim}

The entropy of a discrete random variable $X$
with probability mass function $P$ on the alphabet $A$
is $H(X)=E[-\log P(X)]=\sum_{x\in A}P(x)\log(1/P(x))$.
Throughout the paper, $\log$ denotes the natural logarithm
$\log_e$,
and the support (or alphabet) of any discrete random variable $X$ is
assumed to be a (finite or countably infinite) subset
of the real line or of an arbitrary
discrete abelian group.
Perhaps the simplest bound on the entropy $H(X+Y)$ of the sum
of two independent random variables $X,Y$ is,
$$ H(X+Y)\geq \max\{H(X),H(Y)\},$$
which easily follows from 
elementary properties \cite{cover:book},
\be
H(X)+H(Y)=H(X,Y)=H(Y,X+Y)
&=&H(X+Y)+H(Y|X+Y)
\nonumber\\
&\leq& H(X+Y)+H(Y),
\label{eq:trivial1}
\ee
and similarly with the roles of $X$ and $Y$ interchanged.
The first and third equalities
follow from the chain rule and independence, the
second equality follows from the 
``data processing'' property that 
$H(F(Z))=H(Z)$ if $F$ is a one-to-one function, and
the inequality follows from the fact 
that conditioning reduces entropy.

A similar argument using the nonnegativity of
conditional entropy \cite{cover:book},
$$H(X)+H(Y)=H(Y,X+Y)=H(X+Y)+H(Y|X+Y)\geq H(X+Y),$$
gives the upper bound,
\be
H(X+Y)\leq H(X)+H(Y).
\label{eq:trivial2}
\ee

The starting 
point of our development is the recent
work of Tao \cite{tao:10}, where a series of
sumset bounds are established for $H(X)$,
beginning with the elementary 
inequalities~(\ref{eq:trivial1}) and~(\ref{eq:trivial2}).
The arguments in \cite{tao:10} are largely based 
on the following important observation
\cite{tao:10}\cite{MMT08:pre}:

\begin{lemma}[Functional submodularity of Shannon entropy]
\label{lem:sub}
If $X_0=F(X_1)=G(X_2)$ and $X_{12}=R(X_1,X_2)$,
then:
$$H(X_{12})+H(X_0)\leq H(X_1)+H(X_2).$$
\end{lemma}

\noindent
{\em Proof. }
By data processing for mutual information and
entropy,
$H(X_1)+H(X_2)-H(X_{12})
\geq H(X_1)+H(X_2)-H(X_1,X_2)
= I(X_1;X_2)
\geq I(X_0;X_0)
=H(X_0).$
\qed

One of our main goals in this work is to 
examine the extent to which the
bounds in \cite{tao:10} and in
earlier work extend to the
continuous case.
The differential entropy of a continuous
random variable $X$ with density $f$ on $\RL$
is $h(X)=E[-\log f(X)]=\int_{-\infty}^\infty
f(x)\log(1/f(x))\,dx$. 
The differential entropy of any finite-dimensional,
continuous random vector $\X=(X_1,X_2,\ldots,X_n)$ 
is defined analogously, in terms of the joint
density of the $X_i$.
In order to avoid 
uninteresting technicalities, we assume
throughout that the differential entropies
in the statements of all our results
exist and are finite.

The first important difference between
$H(X)$ and $h(X)$ is that the differential
entropy of a one-to-one function of $X$ is typically
different from that of $X$ itself, even
for linear functions \cite{cover:book}:
For any continuous random vector
$\rm{X}$ and any nonsingular matrix $T$,
$h(T{\rm X})=h({\rm X})+\log|\det(T)|$,
which is different from $h(X)$ unless $T$ 
has determinant equal to $\pm1$.

The upper bound in~(\ref{eq:trivial2}) also 
fails in general for independent continuous $X,Y$:
Take, e.g., $X,Y$ to be independent
Gaussians, one with variance $\sigma^2>2\pi e$
and the other with variance $1/\sigma^2$.
And the functional submodularity Lemma~\ref{lem:sub}
similarly fails for differential entropy. 
For example, taking $X_1=X_2$
an arbitrary continuous random variable with
finite entropy,
$F(x)=G(x)=x$ and $R(x,x')=ax$
for some $a>1$, the obvious differential-entropy analog
of Lemma~\ref{lem:sub} yields $\log a\leq 0$.

On the other hand, the simple lower bound
in (\ref{eq:trivial2}) does generalize,
\be
h(X+Y)\geq \max\{h(X),h(Y)\},
\label{eq:LBtrivial}
\ee
and is equivalent to the data processing
inequality,
$$\min\{I(X+Y;X),\; I(X+Y;Y)\}\geq 0,$$
since,
$$
0
\leq I(X+Y;X)
=h(X+Y)-h(X+Y|X)
=h(X+Y)-h(Y|X)
=h(X+Y)-h(Y),$$
and similarly for $h(X)$ in place of $h(Y)$;
here we use the fact that differential entropy is
translation-invariant.

In the rest of the paper,
all standard properties of $h(X)$ and $H(X)$ will
be used without explicit reference; they can
all be found, e.g., in \cite{cover:book}.
Since it will play a particularly
central role in what follows, we recall
that the mutual information 
$I(\X;\Y)$ between two arbitrary continuous
random vectors $\X,\Y$ can be defined as,
$$I(\X;\Y)
=h(\X)-h(\X|\Y)
=h(\Y)-h(\Y|\X)
=h(\X)+h(\Y)-h(\X,\Y),$$
and the {\em data processing
inequality} states that, whenever 
$\X$ and $\Z$ are conditionally independent
given $\Y$, we have,
$$I(\X;\Y)\geq I(\X,\Z).$$

The development in Section~\ref{s:sumset-h}
will be largely based on the idea that 
the use of functional submodularity can be avoided 
by reducing the inequalities of interest to 
data-processing inequalities for appropriately
defined mutual informations. This reduction is
sometimes straightforward, but sometimes
far from obvious. 

\newpage

\section{Sumset Bounds for Differential Entropy}
\label{s:sumset-h}

Throughout the rest of the paper,
unless explicitly stated otherwise, 
all random variables are assumed to
be real-valued and continuous (i.e., 
with distributions absolutely continuous with respect to
Lebesgue measure, or in other words, having a 
probability density function), and the differential entropy 
of any random variable or random vector appearing 
in the statement of any of our results 
is assumed to exist and be finite.

\subsection{Ruzsa distance and the doubling and difference constants}
\label{s:rddd}

In analogy with the corresponding definition for 
discrete random variables \cite{tao:10}, 
we define the {\em Ruzsa distance} between any
two continuous random variables $X$ and $Y$ as,
$$\dist_R(X,Y)=
h(X'-Y')
-\frac{1}{2}h(X')
-\frac{1}{2}h(Y'),
$$ 
where $X'\sim X$ and $Y'\sim Y$ are
independent. It is obvious that
$\dist_R$ is symmetric, and 
it is nonnegative because of
the lower bound in (\ref{eq:LBtrivial}).
Our first result states that it also satisfies
the triangle inequality:

\begin{theorem} 
\label{thm:Ctriangle}
{\bf (Ruzsa triangle inequality for differential entropy)}
If $X,Y,Z$ are independent, then:
$$h(X-Z)\leq h(X-Y)+h(Y-Z)-h(Y).$$
Equivalently, for 
arbitrary random variables $X,Y,Z$:
$$\dist_R(X,Z)\leq
\dist_R(X,Y)+ \dist_R(Y,Z).$$
\end{theorem}

The proof of the discrete version of this
result in \cite{tao:10}
is based on the discrete entropy
analog of the bound,
\be
h(X,Y,Z)+h(X-Z)\leq h(X-Y,Y-Z)+h(X,Z),
\label{eq:hope}
\ee
which is proved using the functional submodularity
Lemma~\ref{lem:sub}.
Although in general Lemma~\ref{lem:sub}
fails for differential entropy, 
we may try to adapt its proof in this particular setting.
However the obvious modification of the discrete 
proof in the continuous case also fails;
the analog of the first inequality in
the proof of Lemma~\ref{lem:sub}, corresponding 
to $H(X_{12})\leq H(X_1,X_2)$,
is,
$$
h(X,Y,Z)
\leq 
h(X-Y,Y-Z,X,Z),
$$
which is false, since $(X-Y,Y-Z,X,Z)$ is concentrated on a
lower-dimensional subspace of $\RL^4$, and so 
the term on the right side is $-\infty$.
Nevertheless, the actual inequality (\ref{eq:hope})
does hold true.

\begin{lemma}
\label{lem:sub-diff}
The inequality {\em (\ref{eq:hope})} holds true for
any three independent random variables
$X,Y,Z$, and it is equivalent to the 
following data processing inequality:
\be
I(X;(X-Y,Y-Z))
\geq
	I(X;X-Z).
\label{eq:sub-diff}
\ee
\end{lemma}


\noindent
{\em Proof. }
Inequality~(\ref{eq:sub-diff}) is an immediate consequence of
data processing, since $X-Z= (X-Y)+(Y-Z)$, therefore, 
$X$ and $X-Z$ are conditionally independent given $(X-Y,Y-Z)$.
To see that it is equivalent to~(\ref{eq:hope}),
note that the right-hand side of~(\ref{eq:sub-diff}) is,
$$h(X-Z)-h(X-Z|X)=h(X-Z)-h(Z),$$
while the left-hand side is,
\ben
h(X-Y,Y-Z)+h(X)-h(X,X-Y,Y-Z)
&=&
	h(X-Y,Y-Z)+h(X)-h(X,Y,Z)\\
&=&
	h(X-Y,Y-Z)-h(Y)-h(Z),
\een
where the first equality above
follows from the fact that the linear map,
$(x,x-y,y-z)\mapsto (x,y,z)$ has determinant~$1$.
\qed


\noindent
{\em Proof of Theorem~\ref{thm:Ctriangle}. }
Rearranging (\ref{eq:hope}) and using independence,
$$h(X-Z)\leq h(X-Y,Y-Z)-h(Y)
\leq h(X-Y)+h(Y-Z)-h(Y).$$
This is easily seen to be the same as the claimed inequality
upon substituting the definition of the Ruzsa distances in
terms of differential entropies.
\qed


Replacing $Y$ by $-Y$, the triangle inequality yields:

\begin{lemma}
\label{lem:Csumdiff}
If $X,Y,Z$ are independent, then:
$$h(X-Z)+h(Y)\leq h(X+Y)+h(Y+Z).$$
\end{lemma}

In a similar vein we also have:

\begin{lemma}
\label{lem:C3122}
If $X,Y,Z$ are independent, then,
$$h(X+Y+Z)+h(Y)\leq h(X+Y)+h(Y+Z),$$
which is equivalent to the data processing
inequality,
\be
I(X+Y+Z;X)\leq I(X+Y;X).
\label{eq:C3122}
\ee
\end{lemma}

\noindent
{\em Proof. } 
The equivalence of the two stated inequalities follows from the 
observation that 
\ben\begin{split}
I(X+Y+Z;X)
&= h(X+Y+Z)-h(X+Y+Z|X)\\
&=h(X+Y+Z)-h(Y+Z|X)\\
&=h(X+Y+Z)-h(Y+Z),
\end{split}\een
and similarly,
$$
I(X+Y;X) 
=h(X+Y)-h(X+Y|X)
=h(X+Y)-h(Y|X)
=h(X+Y)-h(Y).$$

By the data processing inequality
for mutual information, and the assumed independence,
\ben
	I(X+Y+Z;X)
\leq I(X+Y, Z; X)
= I(X+Y;X) + I(Z;X|X+Y)
=	I(X+Y;X)
\een
which proves~(\ref{eq:C3122}) and hence the lemma.
\qed

Combining the last two lemmas, yields:

\begin{theorem}[Doubling-difference inequality]
\label{thm:Cdoublediff}
If $X_1,X_2$ are i.i.d., then:
$$
\frac{1}{2}
\leq
\frac{h(X_1+X_2)-h(X_1)}{h(X_1-X_2)-h(X_1)}
\leq 2.
$$
Equivalently:
$$
\frac{1}{2}
\leq
\frac{I(X_1+X_2;X_2)}{I(X_1-X_2;X_2)}
\leq 2.
$$
\end{theorem}


\noindent
{\em Proof. } 
For the upper bound, from Lemma~\ref{lem:C3122}
taking $X,-Y$ and $Z$ i.i.d.,
we have,
$$
h(X+Z)+h(Y)
\leq
h(X+Y+Z)+h(Y)
\leq 
h(X+Y)+h(Z+Y),$$
so that,
$$h(X+Z)+h(X)\leq 2h(X-Z),$$
or,
$$h(X+Z)-h(X)\leq 2[h(X-Z)-h(X)],$$
as required.
For the lower bound, Lemma~\ref{lem:Csumdiff} with 
$X,Y,Z$ i.i.d.\ yields,
$$h(X-Y)+h(X)\leq 2h(X+Y),$$
i.e.,
$$h(X-Y)-h(X)\leq 2[h(X+Y)-h(X)],$$
which is the stated lower bound.
The fact that the entropy bounds are equivalent
to the corresponding mutual information bounds
can be established easily as in the first part of 
the proof of~Lemma~\ref{lem:C3122}.
\qed


Theorem~\ref{thm:Cdoublediff} examines a basic constraint that
the differential entropies of the sum and difference of
two i.i.d.\ random variables place on each other.
These quantities are, of course, identical when the density under consideration
is symmetric, but there does not seem to be an immediate intuitive reason
for them to be mutually constraining 
in the case when the difference $X_1-X_2$ has a 
symmetric density but the sum $X_1+X_2$ does not. 
Indeed, Lapidoth and Pete \cite{lapidoth-pete} showed that
the entropies of the sum and difference of two i.i.d.\
random variables can differ by an arbitrarily large amount:
Given any $M>0$, there exist i.i.d.\ $X_1,X_2$
of finite differential entropy,
such that,
\be
h(X_1+X_2)- h(X_1-X_2)>M.
\label{eq:LP}
\ee
If we consider the ``entropy-increase'' due to
addition of subtraction,
\ben
\Delta_{+}&=& h(Y+Y')-h(Y) \\
\Delta_{-}&=& h(Y-Y')-h(Y);
\een
then (\ref{eq:LP}) states that 
the {\it difference} 
$\Delta_{+}-\Delta_{-}$ can be arbitrarily large,
while Theorem~\ref{thm:Cdoublediff} asserts that 
the {\it ratio} $\Delta_{+}/\Delta_{-}$
must always lie between $\half$ and 2.

In other words, we define the {\em doubling constant} and 
the {\em difference constant} of a random variable $X$
as,
$$
\sigma[X]=\exp\{h(X+X')-h(X)\}
\;\;\;\;
\mbox{and}
\;\;\;\;
\delta[X]=\exp\{h(X-X')-h(X)\},
$$
respectively, where $X'$ is an independent
copy of $X$, then Theorem~\ref{thm:Cdoublediff} says that:

\begin{corollary}
\label{cor:Cdoublediff}
For any random variable $X$,
$$\frac{1}{2}\dist_R(X,X)\leq \dist_R(X,-X)\leq 2\dist_R(X,X),$$
equivalently,
$$\delta[X]^{1/2}\leq \sigma[X]\leq \delta[X]^2.$$
\end{corollary}

\noindent
{\em Note. } As mentioned on pp.~64-65 of 
\cite{TV06:book},
the analog of the above upper bound,
$\sigma[X]\leq \delta[X]^2$, in
additive combinatorics is established via 
an application of the 
Pl\"{u}nnecke-Ruzsa inequalities. It is interesting
to note that the entropy version of this result
(both in the discrete and continuous case) can be
deduced directly from elementary arguments.
Perhaps this is less surprising in view of the
fact that strong versions of the
Pl\"{u}nnecke-Ruzsa inequality can also
be established by elementary methods
in the entropy setting, and also because 
of the (surprising and very recent)
work of Petridis \cite{petridis:pre}, 
where an elementary proof of 
the Pl\"{u}nnecke-Ruzsa inequality for sumsets 
is given. See 
Sections~\ref{s:PRI} and~\ref{s:discrete},
and the discussion in \cite{tao-blog-PRI, MMT08:pre}.

We now come to the first result whose proof 
in the continuous case is necessarily significantly 
different than its discrete counterpart.

\begin{theorem}
\label{thm:sum-diff}
{\bf (Sum-difference inequality for differential entropy)}
For any two independent random variables $X,Y$:
\be
h(X+Y)\leq 3h(X-Y)-h(X)-h(Y).
\label{eq:sum-diff2}
\ee
Equivalently, for any pair 
of random variables $X,Y$,
\be
\dist_R(X,-Y)\leq 3\dist_R(X,Y).
\label{eq:sum-diff1}
\ee
\end{theorem}

The equivalence of~(\ref{eq:sum-diff1}) and~(\ref{eq:sum-diff2})
follows simply from the definition of the Ruzsa distance.
Before giving the proof, we state and prove the following
simple version of the theorem in terms of mutual information:

\begin{corollary}[Sum-difference inequality for information]
\label{cor:sum-diff}
For any pair of independent random variables $X,Y$,
and all $0\leq\alpha\leq 1$:
$$
\alpha I(X+Y;X) + (1-\alpha)I(X+Y;Y)
\leq (1+\alpha)I(X-Y;X)+(1+(1-\alpha))I(X-Y;Y).$$
\end{corollary}

\noindent
{\em Proof. }
Subtracting $h(X)$ from both sides of~(\ref{eq:sum-diff2}) yields
$$h(X+Y)-h(X)\leq 3h(X-Y)-2h(X)-h(Y),$$
or equivalently,
$$h(X+Y)-h(X+Y|Y)\leq 2[h(X-Y)-h(X-Y|Y)]+[h(X-Y)-h(X-Y|X)],$$
which, in terms of mutual information becomes,
\be
I(X+Y;Y)\leq 2I(X-Y;Y)+I(X-Y;X).
\label{eq:sd1}
\ee
Repeating the same argument, this time subtracting
$h(Y)$ instead of $h(X)$ from both sides, gives,
\be
I(X+Y;X)\leq 2I(X-Y;X)+I(X-Y;Y).
\label{eq:sd2}
\ee
Multiplying~(\ref{eq:sd1}) by $\alpha$,
(\ref{eq:sd2}) by $(1-\alpha)$, and adding the
two inequalities gives the stated result.
\qed

The inequality~(\ref{eq:sum-diff2}) of 
Theorem~\ref{thm:sum-diff} is an immediate consequence of
the following proposition.

\begin{proposition}
\label{prop:Ccond}
Suppose $X,Y$ are independent, let $Z=X-Y$,
and let $(X_1,Y_1)$ and $(X_2,Y_2)$
be two conditionally independent versions
of $(X,Y)$ given $Z$. If $(X_3,Y_3)\sim(X,Y)$
are independent of $(X_1,Y_1,X_2,Y_2)$,
then:
\be
h(X_3+Y_3)+h(X_1)+h(Y_2)\leq 
h(X_3-Y_2)
+
h(X_1-Y_3)
+
h(Z).
\label{eq:Ccond}
\ee
\end{proposition}

The proof of the discrete analog of the bound~(\ref{eq:Ccond})
in \cite{tao:10} contains two important steps, 
both of which fail for differential entropy.
First, functional submodularity is used 
to deduce the discrete version of,
\be
h(X_1,X_2,X_3,Y_1,Y_2,Y_3) + h(X_3+Y_3)
\leq
	h(X_3,Y_3) 
	+h(X_3-Y_2,X_1-Y_3,X_2,Y_1),
\label{eq:useless}
\ee
but (\ref{eq:useless}) is trivial because
the first term above is equal 
to $-\infty$. Second, the following simple
mutual information identity (implicit in \cite{tao:10})
fails: If $Z=F(X)$ and $X,X'$ are conditionally independent
versions of $X$ given $Z$, then $I(X;X')=H(Z)$. Instead,
for continuous random variables, 
$Z$ and $X$ are conditionally independent
given $X'$, and hence,
$$I(X;X')\geq I(X;Z)= h(Z)-h(Z|X)=+\infty.$$
Instead of this, we will use:

\begin{lemma}
\label{lem:MIC}
Under the assumptions of Proposition~\ref{prop:Ccond}:
$$h(Z,Y_1,Y_2)+h(Z)-h(Y_1)-h(Y_2)=h(X_1)+h(X_2).$$
\end{lemma}

\noindent
{\em Proof. } Expanding and using elementary
properties,
\ben
h(Z,Y_1,Y_2)+h(Z)-h(Y_1)-h(Y_2)
&=&
	h(Y_1,Y_2|Z)+2h(Z)-h(Y_1)-h(Y_2)\\
&=&
	h(Y_1|Z)+h(Y_2|Z)+2h(Z)-h(Y_1)-h(Y_2)\\
&=&
	h(Y_1,Z)+h(Y_2,Z)-h(Y_1)-h(Y_2)\\
&=&
	h(Z|Y_1)+h(Z|Y_2)\\
&=&
	h(X_1-Y_1|Y_1)+h(X_2-Y_2|Y_2)\\
&=&
	h(X_1)+h(X_2),
\een
as claimed.
\qed


\noindent
{\em Proof of Proposition~\ref{prop:Ccond}. }
The most important step of the proof is the
realization that the (trivial) 
result~(\ref{eq:useless}) 
needs to be replaced by the following:
\be
h(Z,X_3,Y_1,Y_2,Y_3) + h(X_3+Y_3)
\leq
	h(X_3,Y_3) 
	+h(X_3-Y_2,X_1-Y_3,X_2,Y_1).
\label{eq:useful}
\ee
Before establishing (\ref{eq:useful}) we note
that it implies,
\ben
h(X_3+Y_3)
\leq
	h(X_3-Y_2)+h(X_1-Y_3)+h(X_2)+h(Y_1)
	-h(Z,Y_1,Y_2),
\een
using the independence of $(X_3,Y_3)$ and $(Y_1,Y_2,Z)$.
Combined with Lemma~\ref{lem:MIC}, this gives the
required result.

To establish (\ref{eq:useful}) we first note
that, by construction, $X_1-Y_1=X_2-Y_2=Z$,
therefore,
\ben
X_3+Y_3
&=&
	X_3+Y_3+(X_2-Y_2)-(X_1-Y_1)\\
&=&
	(X_3-Y_2)-(X_1-Y_3)+X_2+Y_1,
\een
and hence,
by data processing for mutual information,
$$I(X_3;X_3+Y_3)\leq
I(X_3; X_3-Y_2,X_1-Y_3,X_2,Y_1),$$
or, equivalently,
\ben
h(X_3+Y_3)-h(Y_3)
\!&=&\!
	h(X_3+Y_3)-h(X_3+Y_3|X_3)\\
\!&\leq&\!
	h(X_3)+ h(X_3-Y_2,X_1-Y_3,X_2,Y_1)
	-h(X_3-Y_2,X_1-Y_3,X_2,Y_1,X_3)\\
\!&=&\!
	h(X_3)+ h(X_3-Y_2,X_1-Y_3,X_2,Y_1)
	-h(Z,Y_1,Y_2,Y_3,X_3),
\een
where the last equality follows from the fact 
that the linear map,
$(z,y_1,y_2,y_3,x_3)\mapsto(x_3-y_2,y_1+z-y_3,y_2+z,y_1,x_3)$,
has determinant~$1$. Rearranging and using the 
independence of $X_3$ and $Y_3$ gives (\ref{eq:useful}) 
and completes the proof.
\qed

\subsection{The differential entropy Pl\"{u}nnecke-Ruzsa inequality}
\label{s:PRI}

In additive combinatorics, the 
Pl\"{u}nnecke-Ruzsa inequality
for iterated sumsets
is a subtle result that was originally
established through an involved
proof based on the theory of commutative 
directed graphs;
see Chapter~6 of \cite{TV06:book}. 
It is interesting that
its entropy version can be proved as
a simple consequence of the data
processing bound in Lemma~\ref{lem:C3122}.
See also the remark following 
Corollary~\ref{cor:Cdoublediff}
above.

\begin{theorem}[Pl\"{u}nnecke-Ruzsa inequality for differential entropy]
\label{thm:CPR}
Suppose that the random variables 
$X,Y_1,$ $Y_2,$ $\ldots,Y_n$ are independent,
and that, for each $i$, $Y_i$ is only weakly dependent on $(X+Y_i)$, 
in that $I(X+Y_i;Y_i)\leq \log K_i$ for 
finite constants $K_1,K_2,\ldots,K_n$. In other words,
$$h(X+Y_i)\leq h(X) + \log K_i,\;\;\;\;\mbox{for each }i.$$
Then,
$$h(X+Y_1+Y_2+\cdots+Y_n)\leq h(X)+\log K_1K_2\cdots K_n,$$
or, equivalently,
$$I(X+Y_1+Y_2+\cdots+Y_n; Y_1+Y_2+\cdots+Y_n)
\leq \log K_1K_2\cdots K_n.$$
\end{theorem}

\noindent
{\em Proof. }
Using Lemma~\ref{lem:C3122}:
\ben
h(X+Y_1+Y_2+\cdots+Y_n)
&\leq&
	h(X+Y_1+Y_2+\cdots+Y_{n-1})
	+h(X+Y_n)-h(X)\\
&\leq&
	h(X+Y_1+Y_2+\cdots+Y_{n-1})
	+\log K_n,
\een
and continuing inductively yields
the result.
\qed


By an application
of the entropy Pl\"{u}nnecke-Ruzsa 
inequality we can establish the following
bound on iterated sums.

\begin{theorem}[Iterated sum bound]
\label{thm:iterated}
Suppose $X,Y$ are independent
random variables,
let $(X_0,Y_0), (X_1,Y_1),\ldots,(X_n,Y_n)$ be i.i.d.\
copies of $(X,Y)$, and write $S_i=X_i+Y_i$ for
the sums of the pairs, $i=0,1,\ldots,n$. Then:
$$h(S_0+S_1+\cdots+S_n)\leq(2n+1)h(X+Y)-nh(X)-nh(Y).$$
\end{theorem}

{\em Proof. }
Suppose the result is true for $n=1$. Then,
for each $i$,
$$h(S_0+S_i)\leq 3h(X+Y)-h(X)-h(Y)=h(S_0)+[2h(X+Y)-h(Y)-h(Y)],$$
and the general claim follows
from an application of the 
entropy Pl\"{u}nnecke-Ruzsa inequality
(Theorem~\ref{thm:CPR}). The case $n=1$ 
is an immediate consequence of the following lemma
(which generalizes Lemma~\ref{lem:C3122}) with 
$X\sim Z$ and $Y\sim W$.
\qed

\begin{lemma}
If $X,Y,Z,W$ are independent, then:
$$h(X+Y+Z+W)+h(Y)+h(Z)\leq h(X+Y)+h(Y+Z)+h(Z+W).$$
\end{lemma}

\noindent
{\em Proof. } 
Applying Lemma~\ref{lem:C3122} with $Z+W$ in place of $Z$,
$$h(X+Y+Z+W)+h(Y)\leq h(X+Y)+h(Y+Z+W),$$
and using Lemma~\ref{lem:C3122} again on the last term above,
$$h(X+Y+Z+W)+h(Y)\leq h(X+Y)+h(Y+Z)+h(Z+W)-h(Z).$$
Rearranging, proves the claim.
\qed

Let us briefly comment on the interpretation of Theorem~\ref{thm:iterated}.
The result may be rewritten as
\ben
(n+1) h(X+Y) - h\bigg(\sum_{i=1}^n X_i + \sum_{i=1}^n Y_i\bigg)
\geq n [h(X)+h(Y)-h(X+Y)] ,
\een
and hence as
\be\label{eq:it-sum}
h(X_0+Y_0, \ldots, X_n+Y_n)  - h\bigg(\sum_{i=1}^n X_i + \sum_{i=1}^n Y_i\bigg)
\geq n [h(X,Y)-h(X+Y)] .
\ee
Thus the ``differential entropy loss from summation'' of the collection of $n+1$
independent random variables $\{X_i+Y_i: i=1,\ldots, n\}$ is at least $n$ times
the ``differential entropy loss from summation'' of the two independent random variables $\{X,Y\}$.
(In the discrete case, one would have a stronger interpretation as the entropy loss
would be precisely the {\it information} lost in addition.)

\subsection{The differential entropy Balog-Szemer\'{e}di-Gowers lemma} 

The differential entropy version of the 
{\em Balog-Szemer\'{e}di-Gowers lemma}
stated next says that,
if $X,Y$ are weakly dependent and $X+Y$ has small entropy, 
then there exist conditionally independent 
versions of $X,Y$ that have almost the same entropy, and 
whose {\em independent} sum still has small entropy.

\begin{theorem}[Balog-Szemer\'{e}di-Gowers lemma for differential entropy]
\label{thm:CBSG}
Suppose $X$, $Y$ are
weakly dependent in the sense that $I(X;Y)\leq \log K$, i.e.,
\begin{equation}
	\label{Centk}
 h(X,Y) \geq h(X) + h(Y) - \log K,
 \end{equation}
for some $K \geq 1$, and suppose also that, 
\begin{equation}\label{Centk2}
 h(X+Y) \leq \frac{1}{2} h(X) + \frac{1}{2} h(Y) + \log K.
 \end{equation}
Let $X_1,X_2$ be conditionally independent versions of $X$ given $Y$,
and let $Y'$ be a conditionally independent version of $Y$,
given $X_2$ and $Y$; in other words, the sequence
$X_2,Y,X_1,Y'$ forms a Markov chain.
Then:
\begin{align}
h( X_2 | X_1, Y ) 	&\geq h(X) - \log K \label{Cloga}\\
h( Y' | X_1, Y ) 	&\geq h(Y) - \log K \label{Clogb}\\
h( X_2 + Y' | X_1, Y )  &\leq \frac{1}{2} h(X) 
+ \frac{1}{2} h(Y) + 7 \log K.\label{Cseven}
\end{align}
\end{theorem}

Following the corresponding development
in \cite{tao:10} for discrete random variables,
first we establish a weaker result in the following
proposition. The main step in the proof -- which is
also a very significant 
difference from the proof of the discrete
version of the result in \cite{tao:10} -- is the
identification of the ``correct'' data processing
bound~(\ref{eq:dpkey1}) that needs to replace
the use of functional submodularity.

\begin{proposition}
\label{prop:Cweak}
{\bf (Weak Balog-Szemer\'edi-Gowers lemma)}
Under the assumptions of Theorem~\ref{thm:CBSG},
we have:
$$ h(X_1-X_2|Y) \leq h(X) + 4 \log K.$$
\end{proposition}

\noindent
{\em Proof. }
Let $X_1$ and $X_2$ be conditionally independent as above, 
and let $(X_1,Y,X_2)$ and 
$(X_1,Y'',X_2)$ be conditionally independent
versions of 
$(X_1,Y,X_2)$, given $(X_1,X_2)$.
We claim that,
\be
h( X_1,X_2,Y,Y'' ) + h(X_1-X_2, Y) 
\leq 
h(X_1,X_2,Y) + h(X_1+Y'', X_2+Y'',Y),
\label{eq:Csub}
\ee
which is equivalent to,
\be
h( X_1,X_2,Y''|Y) + h(X_1-X_2|Y) 
\leq 
h(X_1,X_2|Y) + h(X_1+Y'', X_2+Y''|Y).
\label{eq:Csub2}
\ee
This follows from the data processing argument:
\be
&&
h(X_1-X_2|Y)\nonumber\\
&&=
	I(X_1-X_2; X_1|Y)
	+h(X_1-X_2|X_1,Y)
	\nonumber\\
&&\leq
	I(X_1+Y'', X_2+Y''; X_1|Y)
	+h(X_2|Y)
	\label{eq:dpkey1}\\
&&=
	h(X_1+Y'', X_2+Y''|Y)
	+h(X_1|Y)
	-h(X_1+Y'', X_2+Y'', X_1|Y)
	+h(X_2|Y)\nonumber\\
&&=
	h(X_1,X_2|Y)
	+h(X_1+Y'', X_2+Y''|Y)
	-h(X_1+Y'', X_2+Y'', X_1|Y)
	\nonumber\\
&&=
	h(X_1,X_2|Y)
	+h(X_1+Y'', X_2+Y''|Y)
	-h(X_1,X_2,Y''|Y),
	\nonumber
\ee
where the last equality follows from the fact
that the linear map $(x_1,x_2,y)\mapsto(x_1+y,x_2+y,x_1)$
has determinant~$-1$. This 
establishes~(\ref{eq:Csub2})
and hence~(\ref{eq:Csub}).

We now deduce the result from~(\ref{eq:Csub}).
By the independence bound for joint entropy,
the second term in the right-hand side of (\ref{eq:Csub}) is,
$$h(X_1+Y'', X_2+Y'',Y)\leq
2h(X+Y) + h(Y).$$
By conditional independence and the chain rule,
the first term in the right-hand side of (\ref{eq:Csub}) is,
$$
h(X_1,X_2,Y)=
h(X_1,X_2|Y)+h(Y)=
h(X_1|Y)+
h(X_2|Y)+h(Y)=
2h(X,Y)-h(Y).$$
Using this, conditional independence, and the independence
bound for joint entropy,
the first term in the left-hand of (\ref{eq:Csub}) is,
\ben
h( X_1,X_2,Y,Y'' ) 
&=& h(X_1,X_2)+h(Y,Y''|X_1,X_2)\\
&=& h( X_1,X_2) +h(Y|X_1,X_2)+h(Y''|X_1,X_2)\\
&=& 2h(X_1,X_2,Y)-h(X_1,X_2)\\
&=& 4h(X,Y)-2h(Y)-h(X_1,X_2)\\
&\geq& 4h(X,Y)-2h(Y)-2h(X).
\een
And by the chain rule, the second term in the
left-hand side of (\ref{eq:Csub}) is,
$$h(X_1-X_2,Y)
=
	h(X_1-X_2|Y)+h(Y).$$
Finally combining all the above estimates yields,
$$h(X_1-X_2|Y)+h(Y)
+ 4h(X,Y)-2h(Y)-2h(X)
\leq
2h(X+Y) + h(Y)
+2h(X,Y)-h(Y),$$
or,
$$h(X_1-X_2|Y)
\leq
2h(X+Y) 
+h(Y)
- 2h(X,Y)+2h(X),$$
and the claim then follows from 
the assumptions in \eqref{Centk}, \eqref{Centk2}.
\qed

The proof of Theorem~\ref{thm:CBSG}, given next,
is similar. Again, the key step is an application
of the data processing inequality in (\ref{eq:dpkey2}).

\medskip

\noindent
{\em Proof of Theorem~\ref{thm:CBSG}.  }
The bound \eqref{Cloga} immediately follows from \eqref{Centk}
and the definitions:
$$h(X_2|X_1,Y)=h(X_2|Y)=h(X|Y)=h(X,Y)-h(Y)\geq h(X)-\log K.$$
Similarly, \eqref{Clogb} follows from \eqref{Centk}:
$$h(Y'|X_1,Y)=h(Y'|X_1)=h(Y|X)=h(X,Y)-h(X)\geq h(Y)-\log K.$$
For \eqref{Cseven}, we first claim that the following holds,
\be
h(X_1,X_2,Y,Y') + h(X_2+Y',Y) \leq h(X_2, Y', Y) 
+ h(X_1-X_2,X_1+Y', Y),
\label{eq:bsg}
\ee
or, equivalently,
\be
h(X_1,X_2,Y'|Y) + h(X_2+Y'|Y) \leq h(X_2, Y'| Y) 
+ h(X_1-X_2,X_1+Y'|Y).
\label{eq:bsg2}
\ee
As in the previous proof, this follows from a data-processing
argument,
\be
h(X_2+Y'|Y)
&=&
	I(X_2+Y';X_2|Y)+h(X_2+Y'|X_2,Y)
	\nonumber\\
&\leq&
	I(X_1-X_2,X_1+Y';X_2|Y)+h(Y'|X_2,Y)
	\label{eq:dpkey2}\\
&=&
	h(X_1-X_2,X_1+Y'|Y)
	+h(X_2|Y)
	-h(X_1-X_2,X_1+Y',X_2|Y)
	+h(Y'|Y)
	\nonumber\\
&=&
	h(X_1-X_2,X_1+Y'|Y)
	-h(X_1,X_2,Y'|Y)
	+h(X_2,Y'|Y),
	\nonumber
\ee
where the last equality follows from the fact that 
$X_2$ and $Y'$ are conditionally independent given $Y$,
and also from the fact that the linear map
$(x_1,x_2,y)\mapsto(x_1-x_2,x_1+y,x_2)$
has determinant~$-1$. This proves~(\ref{eq:bsg2})
and hence~(\ref{eq:bsg}).

As in the previous proof, we bound each of
the four terms in~(\ref{eq:bsg}) as follows.
By the chain rule and conditional independence,
the first term is,
\begin{align*}
h(X_1,X_2,Y,Y') &= h(X_1,X_2,Y) + h(Y'|X_1,X_2,Y) \\
&= h(X_1,X_2,Y) + h(Y'|X_1) \\
&= h(Y)+2h(X|Y) + h(X,Y) -h(X) \\
&= 3h(X,Y) - h(X) - h(Y).
\end{align*}
By the chain rule the second term is,
$h(X_2+Y',Y) = h(X_2+Y'|Y) + h(Y)$,
and by the independence entropy bound
for the third term we have,
$$h(X_2,Y',Y) \leq h(X_2,Y) + h(Y') = h(X,Y) + h(Y).$$
Finally by the chain rule and the fact that 
conditioning reduces entropy,
\begin{align*}
h(X_1-X_2,X_1+Y',Y) &\leq h(Y)+h(X_1-X_2|Y) + h(X_1+Y') \\
&= h(X_1-X_2|Y) + h(Y) + h(X+Y).
\end{align*}
Substituting these into (\ref{eq:bsg}) gives,
$$ h(X_2+Y'|Y) \leq h(X_1-X_2|Y) + 2h(Y) 
+ h(X) + h(X+Y) - 2h(X,Y),$$
and applying the weak 
Balog-Szemer\'{e}di-Gowers lemma
of Proposition~\ref{prop:Cweak} 
together with \eqref{Centk} and \eqref{Centk2}, yields,
$$ 
h(X_2+Y'|X_1,Y) =
h(X_2+Y'|Y) 
\leq \frac{1}{2} h(X) + \frac{1}{2} h(Y) + 7 \log K,$$
as claimed.
\qed

Let us comment some more on the interpretation of Theorem~\ref{thm:CBSG}.
The conditions assumed may be rewritten as follows:
\begin{enumerate}
\item 
The variables $(X_2,Y,X_1,Y')$ form a Markov chain,
with the marginal distributions of the 
pairs $(X_2, Y), (X_1, Y)$ and $(X_1, Y')$
all being the same as the distribution of $(X,Y)$.
\item $I(X;Y)\leq \log K$, i.e., if we represent the 
Markov chain $(X_2,Y,X_1,Y')$
on a graph, then the information shared across each 
edge is the same (by 1.)
and it is bounded by $\log K$.
\item $I(X+Y;X)+I(X+Y;Y)\leq 2\log K$.
\end{enumerate}
The first 2 parts of the conclusion may be rewritten as:
\begin{enumerate}
\item $I( X_2 ; X_1, Y ) \leq \log K$;
\item $I( Y' ; X_1, Y )\leq \log K$.
\end{enumerate}
These are obvious from looking at the graph structure of the
dependence. To rewrite the third part of the conclusion, note that,
\ben\begin{split}
h(X)+h(Y)-2h( X_2 + Y' | X_1, Y )  
&= [h(X_2)-h(X_2| X_1, Y)] + [h(Y')-h(Y'| X_1, Y)] \\
&\quad + [h(X_2| X_1, Y)- h( X_2 + Y' | X_1, Y )] \\
&\quad + [h(Y'| X_1, Y)- h( X_2 + Y' | X_1, Y )]\\
&= I( X_2 ; X_1, Y ) + I( Y' ; X_1, Y ) \\
&\quad + I(X_2+Y';Y'|X_1, Y) 
+ I(X_2+Y';X_2|X_1, Y)  ,
\end{split}\een
so that using the first 2 parts of the conclusion, the third part
says that
\ben
I(X_2+Y';Y'|X_1, Y) + I(X_2+Y';X_2|X_1, Y) 
\leq 16 \log K.
\een
This is not the same as saying that the boundedness of $I(X+Y;X)+I(X+Y;Y)$
for the dependent pair $(X,Y)$ translates to boundedness of the corresponding
quantity for independent $X$ and $Y$ with the same marginals (since
conditioning will change the marginal distributions), but it does mean
that if we embed the dependent pair $(X_1,Y)$ into a Markov chain
that has $X_2$ and $Y'$ at the ends, one has boundedness {\it on average} of 
the corresponding {\it conditional} quantity for the pair $(X_2,Y')$
(which is conditionally independent given $X_1$ and $Y$).

\subsection{Sumset bounds for discrete entropy}
\label{s:discrete}

Here we give a brief discussion of 
the discrete versions of the results 
presented so far in this section, 
their origin and the corresponding 
discrete proofs.

The discrete version of the Ruzsa 
triangle inequality as in 
Theorem~\ref{thm:Ctriangle} 
was given in \cite{Ruz09} 
and \cite{tao:10}. 
The analog of Lemma~\ref{lem:sub-diff} 
for discrete random variables was
established in \cite{tao:10},
and of Lemma~\ref{lem:C3122}
in \cite{MMT08:pre}.
The discrete entropy version 
of the lower bound in 
the doubling-difference inequality
of Theorem~\ref{thm:Cdoublediff}
is implicit in 
\cite{Ruz09} and \cite{tao:10},
and the corresponding upper bound
is implicitly derived in \cite{MMT08:pre}.
The discrete version of 
the sum-difference inequality of
Theorem~\ref{thm:sum-diff} is proved
in \cite{tao:10}; the form given in
Corollary~\ref{cor:sum-diff} in terms
of mutual information is new even
in the discrete case, 
as is Lemma~\ref{lem:MIC}.

The discrete analog of
Proposition~\ref{prop:Ccond}
is implicit in \cite{tao:10}.
The Pl\"{u}nnecke-Ruzsa inequality
(Theorem~\ref{thm:CPR}) for discrete
random variables in 
implicitly proved in \cite{KV:83},
and explicitly stated and discussed in
\cite{tao-blog-PRI}.
The iterated sum bound of
Theorem~\ref{thm:iterated}
in the discrete case is implicit
in \cite{tao:10}, while
the discrete versions of the strong
and weak forms of the 
Balog-Szemer\'{e}di-Gowers lemma
(Theorem~\ref{thm:CBSG} and
Proposition~\ref{prop:Cweak})
are both given in \cite{tao:10}.

Finally, in the unpublished notes
of Tao and Vu \cite{tao-vu:notes}, 
the following is stated as an exercise:

\begin{proposition}[Ruzsa covering lemma for Shannon entropy]
\label{prop:covering}
Suppose $X,Y$ are independent discrete random
variables, and let $(X_1,Y_1),$ $(X_2,Y_2)$ 
be versions of $(X,Y)$ that are conditionally independent given $X+Y$. Then:
\be
H(X_1,X_2,Y_1|Y_2)=2H(X)+H(Y)-H(X+Y).
\label{eq:rcl}
\ee
\end{proposition}

We give a proof below for the sake of completeness,
but first we note that the result actually
fails for differential entropy: By construction
we have that $Y_2=X_1+Y_1-X_2$, therefore,
the left-hand side of the continuous analog of~(\ref{eq:rcl})
is,
$$h(X_1,X_2,Y_1|Y_2)=h(X_1,X_2,Y_2|Y_2)=-\infty.$$

\noindent
{\em Proof. }
Since, by definition, $X_1+Y_1=X_2+Y_2$, the triplet
$(X_1,X_2,X_1+Y_1)$ determines all four random variables.
Therefore, by data processing for the discrete
entropy and elementary properties, we have,
\ben
H(X_1,X_2,Y_1|Y_2)
&=&
	H(X_1,X_2,Y_1,Y_2)-H(Y_2)\\
&=&
	H(X_1,X_2,X_1+Y_1)-H(Y)\\
&=&
	H(X_1+Y_1)+ H(X_1,X_2|X_1+Y_1)-H(Y)\\
&=&
	H(X+Y)+ H(X_1|X_1+Y_1)+H(X_2|X_1+Y_1)-H(Y)\\
&=&
	2H(X,X+Y)-H(X+Y)-H(Y)\\
&=&
	2H(X,Y)-H(X+Y)-H(Y)\\
&=&
	2H(X)+H(Y)-H(X+Y),
\een
as claimed.
\qed

\newpage

\section{A Differential Entropy Inverse Sumset Theorem}
\label{s:inverse-h}

The inverse sumset theorem of 
Freiman-Green-Ruzsa states that,
if a discrete set is such that
the cardinality of the sumset 
$A+A$ is close to the cardinality
of $A$ itself, then $A$ is
``as structured as possible'' in that
it is close to a generalized arithmetic
progression; see \cite{green-ruzsa:07} or
\cite{TV06:book} for details.
Roughly speaking, the discrete entropy
version of this result established by
Tao \cite{tao:10} says that, if $X,X'$
are i.i.d.\ copies of a discrete random
variable and $H(X+X')$ is not much larger
than $H(X)$, then the distribution of 
$X$ is close to the uniform distribution
on a generalized arithmetic progression.
In other words, if the doubling constant
$\sigma[X]=\exp\{H(X+X')-H(X)\}$ is small,
then $X$ is close to having a maximum
entropy distribution.

Here we give a quantitative version
of this result for continuous random
variables. First we note that the
entropy power inequality \cite{cover:book}
for i.i.d.\ summands states that,
$$e^{2h(X+X')}\geq 2e^{2h(X)},$$
or, equivalently, recalling
the definition of the doubling constant
from Section~\ref{s:rddd},
$$\sigma[X]:=\exp\{h(X'+X)-h(X)\}\geq \sqrt{2},$$
with equality iff $X$ is Gaussian.
Note that, again, the extreme case
where $h(X+X')$ is as close as possible
to $h(X)$ is attained by the distribution
which has maximum entropy on $\RL$, 
subject to a variance constraint.

Next we give conditions under which
the doubling constant $\sigma[X]$ of 
a continuous random variable is small 
only if the distribution of $X$ is 
appropriately close to being Gaussian.
Recall that the {\em Poincar\'{e} constant}
$R(X)$ of a continuous random variable $X$
is defined as,
$$R(X)=\sup_{g\in H_1(X)}\frac{E[g(X)^2]}{E[g'(X)^2]},$$
where the supremum is over all functions $g$
in the space $H_1(X)$ of absolutely continuous
functions with $E[g(X)]=0$ and $0<\VAR(g(X))<\infty$.
As usual, we write $D(f\|g)$ for the relative entropy
$\int f\log(f/g)$ between two densities
$f$ and $g$.

\begin{theorem}
\label{thm:inverse}
{\bf (Freiman-Green-Ruzsa theorem for differential entropy)}
Let $X$ be an arbitrary continuous random variable with density $f$.
\begin{itemize}
\item[{\em (i)}]
$\sigma[X]\geq \sqrt{2}$, with equality iff $X$ is Gaussian.
\item
[{\em (ii)}]
If $\sigma[X]\leq C$ and $X$ has finite Poincar\'{e}
constant $R=R(X)$, then $X$ is approximately Gaussian
in the sense that,
$$
\frac{1}{2}\|f-\phi\|_1^2
\leq D(f\|\phi)\leq 
\Big(\frac{2R}{\sigma^2}+1\Big)\log\Big(\frac{C}{\sqrt{2}}\Big),
$$
where $\sigma^2$ is the variance of $X$
and $\phi$ denotes the Gaussian density with
the same mean and variance as $X$.
\end{itemize}
\end{theorem}

At first sight, the assumption of a finite Poincar\'{e} 
constant may appear unnecessarily restrictive in the
above result. Indeed, we conjecture that this assumption
may be significantly relaxed. On the other hand,
a related counterexample by Bobkov, Chistyakov and G\"{o}tze
\cite{bobkov:pre} suggests that there is good reason for 
caution.

\begin{theorem}
\label{thm:inverse2}
Let $X$ be an arbitrary continuous random variable with density $f$
and finite variance. Let $\phi$ be the Gaussian density with the same 
mean and variance as $f$.
\begin{itemize}
\item[{\em (i)}]
$\sigma[X]\leq \sqrt{2}\,\exp\{D(f\|\phi)\},\;$
with equality iff $X$ is Gaussian,
\item
[{\em (ii)}] 
{\em \cite{bobkov:pre}}
For any $\eta>0$ there exists a continuous
random variable $X$ with,
\be
\sigma[X]>  (\sqrt{2}-\eta) \exp\{D(f\|\phi)\},
\label{eq:sbig}
\ee
but with a distribution well separated
from the Gaussian, in that,
\be
\|f-\phi\|_1>C,
\label{eq:TV}
\ee
where $C$ is an absolute constant
independent of $\eta$.
\end{itemize}
\end{theorem}

\noindent
{\em Proof Theorem~\ref{thm:inverse}. }
Part~(i) follows from the entropy power inequality,
as discussed in the beginning of the section,
and the first inequality in part~(ii) is simply Pinsker's
inequality \cite{cover:book}.

For the main estimate, assume without loss
of generality that $X$ has zero mean, 
and recall that Theorem~1.3
of \cite{johnson-barron:04}
says that, 
$$D\Big(\frac{X+X'}{\sqrt{2}}\Big)\leq D(X)\Big(
\frac{2R}{\sigma^2+2R}\Big),$$
where, for any finite-variance, 
continuous random variable $Y$ with density $g$,
$D(Y)$ denotes the relative entropy
between $g$ and the normal density $\phi_Y$
with the same mean and variance as $Y$.
Since $D(Y)$ can be expanded to,
$D(Y)=h(\phi_Y)-h(Y),$ the above expression simplifies
to, 
\be
h\Big(\frac{X+X'}{\sqrt{2}}\Big)-h(X)\geq
\Big(\frac{\sigma^2}{2R+\sigma^2}\Big) 
[h(\phi)-h(X)],
\label{eq:JB}
\ee
or, 
$$
\log\Big(\frac{C}{\sqrt{2}}\Big)\geq
\log\Big(\frac{\sigma[X]}{\sqrt{2}}\Big)
\geq 
\Big(\frac{\sigma^2}{2R+\sigma^2}\Big) 
D(f\|\phi),$$
as claimed.
\qed

\noindent
{\em Proof of Theorem~\ref{thm:inverse2}. }
As in the last proof, for any 
finite-variance, continuous random variable $Y$ 
with density $g$, write $D(Y)$ for the relative entropy
between $g$ and the normal density $\phi_Y$
with the same mean and variance as $Y$, so that
$D(Y)=h(\phi_Y)-h(Y).$ Then, letting
$X,X'$ be two i.i.d.\ copies of $X$,
\be
D\Big(\frac{X+X'}{\sqrt{2}}\Big)
&=&
	h(\phi)-h\Big(\frac{X+X'}{\sqrt{2}}\Big)
	\nonumber\\
&=&
	h(X)-h(X+X')+
	h(\phi)-h(X)+
	\log \sqrt{2}
	\nonumber\\
&=&
	\log\Big(\frac{\sqrt{2}\exp\{D(f\|\phi)\}}{\sigma[X]}\Big).
	\label{eq:identify}
\ee
The result of part~(i) follows from~(\ref{eq:identify})
upon noting that relative entropy is always nonnegative.
Part~(ii) is a simple restatement of the counterexample
in Theorem~1.1 of \cite{bobkov:pre}. Taking in their
result $\epsilon=-\log(1-\eta/\sqrt{2})$, we are
guaranteed the existence of an absolute constant
$C$ and a random variable $X$ 
such that~(\ref{eq:TV}) holds and $D(X+X')<\epsilon$.
But, using~(\ref{eq:identify}) this translates to,
$$
-\log\Big(1-\frac{\eta}{\sqrt{2}}\Big)=\epsilon>D(X+X')=
D\Big(\frac{X+X'}{\sqrt{2}}\Big)
=	\log\Big(\frac{\sqrt{2}\exp\{D(f\|\phi)\}}{\sigma[X]}\Big),$$
and, rearranging, this is exactly condition~(\ref{eq:sbig}).
The fact that $X$ can be chosen to have finite variance
is a consequence of the remarks following 
Theorem~1.1 in \cite{bobkov:pre}. 
\qed

We close this section by noting that the two results
above actually generalize to the {\em difference 
constant},
$$\delta[X]:=\exp\{h(X-X')-h(X)\},$$
for i.i.d.\ copies $X,X'$ of $X$,
in place of the doubling constant $\sigma[X]$.

\begin{corollary}
\label{cor:inverse3}
Let $X$ be an arbitrary continuous random variable with density $f$.
\begin{itemize}
\item[{\em (i)}]
$\delta[X]\geq \sqrt{2}$, with equality iff $X$ is Gaussian.
\item
[{\em (ii)}]
If $\delta[X]\leq C$ and $X$ has finite Poincar\'{e}
constant $R=R(X)$, then $X$ is approximately Gaussian
in the sense that,
$$
\frac{1}{2}\|f-\phi\|_1^2
\leq D(f\|\phi)\leq 
\Big(\frac{2R}{\sigma^2}+1\Big)\log\Big(\frac{C^2}{\sqrt{2}}\Big),
$$
where $\sigma^2$ is the variance of $X$
and $\phi$ denotes the Gaussian density with
the same mean and variance as $X$.
\end{itemize}
\end{corollary}

\noindent
{\em Proof. }
Since the entropy power inequality \cite{cover:book}
holds for arbitrary independent random variables,
the proof of~(i) is identical to that in
Theorem~\ref{thm:inverse}, with $-X'$ in place of $X'$.
For~(ii)
we assume again without loss 
of generality that $X$ has zero mean 
and recall that, 
from Theorem~\ref{thm:Cdoublediff}, we have,
$$h(X+X')\leq 2h(X-X')-h(X).$$ 
Combining this with the estimate~(\ref{eq:JB})
obtained in the proof of Theorem~\ref{thm:inverse},
yields,
$$2h(X-X')
-2h(X)
-\log\sqrt{2}
\geq
\Big(\frac{\sigma^2}{2R+\sigma^2}\Big) 
[h(\phi)-h(X)].$$
or, 
$$
\log\Big(\frac{C^2}{\sqrt{2}}\Big)\geq
\log\Big(\frac{\delta[X]}{\sqrt{2}}\Big)
\geq 
\Big(\frac{\sigma^2}{2R+\sigma^2}\Big) 
D(f\|\phi),$$
as claimed.
\qed

\begin{corollary}
\label{thm:inverse4}
Let $X$ be an arbitrary continuous random variable with density $f$
and finite variance. Let $\phi$ be the Gaussian density with the same 
mean and variance as $f$.
\begin{itemize}
\item[{\em (i)}]
$\delta[X]\leq \sqrt{2}\,\exp\{D(f\|\phi)\},\;$
with equality iff $X$ is Gaussian,
\item
[{\em (ii)}] 
For any $\eta>0$ there exists a continuous
random variable $X$ with,
\be
\delta[X]>  (\sqrt{2}-\eta) \exp\{D(f\|\phi)\},
\label{eq:sbig2}
\ee
but with a distribution well separated
from the Gaussian, in that,
\be
\|f-\phi\|_1>C,
\label{eq:TV2}
\ee
where $C$ is an absolute constant
independent of $\eta$.
\end{itemize}
\end{corollary}

\noindent
{\em Proof. }
As in the proof of Theorem~\ref{thm:inverse2},
with $-X'$ in place of $X'$ we have,
\ben
0\leq D\Big(\frac{X-X'}{\sqrt{2}}\Big)
&=&
	h(\phi)-h\Big(\frac{X-X'}{\sqrt{2}}\Big)\\
&=&
	h(X)-h(X-X')+
	h(\phi)-h(X)+
	\log \sqrt{2}\\
&=&
	\log\Big(\frac{\sqrt{2}\exp\{D(f\|\phi)\}}{\delta[X]}\Big),
\een
giving~(i).
Part~(ii) follows from
Theorem~1.1 of \cite{bobkov:pre}
exactly as in the proof of the 
corresponding result in 
Theorem~\ref{thm:inverse2},
since the distribution of the
random variables in the counterexample
given in \cite{bobkov:pre} can be taken
to be symmetric, so that $X+X'$ has 
the same distribution as $X-X'$.
\qed

\bibliographystyle{plain}

\def\cprime{$'$}

\end{document}